\documentclass[pre,aps,twocolumn,amsmath,amssymb,longbibliography]{revtex4-1}

\pdfoutput=1
\usepackage{hyperref}
\usepackage{graphicx}
\usepackage[usenames]{color}
\usepackage{bm}
\usepackage[final]{movie15}

\definecolor{blue}{rgb}{0,0,0}

\def\f#1{Fig.~\ref{#1}}

\def\c#1{~\cite{#1}}
\def\cc#1{Ref.\c{#1}}

\def\ta0{\tilde{a}_0}

\def\beq{\begin{equation}}
\def\eeq{\end{equation}}
\def\bea{\begin{eqnarray}}
\def\eea{\end{eqnarray}}

\begin{document}

\title{Comment on the literature definition(s) of ``dynamical phase transition''}
\author{Stephen Whitelam}
\affiliation{Molecular Foundry, Lawrence Berkeley National Laboratory, 1 Cyclotron Road, Berkeley, CA 94720, USA}


\maketitle

\noindent {\em Introduction.} Dynamical phase transitions are often associated with a singular response of the dynamical large-deviation functions of a model, particularly a kink of the scaled cumulant-generating function (SCGF)\c{garrahan2009first,vaikuntanathan2014dynamic,nemoto2017finite,nyawo2017minimal,horowitz2017stochastic,nyawo2018dynamical,nemoto2019optimizing,jack2020ergodicity}. Such kinks are usually said to indicate the existence of a first-order dynamical phase transition and accompanying dynamical phase coexistence\c{garrahan2009first,jack2020ergodicity}, by analogy with equillibrium systems\c{touchette2009large,binney1992theory}. Kinks can emerge in various settings, including the small-rate (reducible) limits of microscopic models\c{dinwoodie1992large,dinwoodie1993identifying,coghi2019large,garrahan2014comment,whitelam2018large}, and the large-size limit of lattice- or off-lattice models\c{garrahan2009first,nemoto2019optimizing,buvca2019exact}. 

There is not universal agreement about when the label ``dynamical phase transition'' applies: some authors consider to the term to be appropriate only in the large-size limit of a many-body system\c{buvca2019exact}; some consider that it also applies in the large-size limit of a one-body system\c{vaikuntanathan2014dynamic,nemoto2017finite}; and some apply the term to systems of finite size\c{nyawo2017minimal,horowitz2017stochastic}. My view is that the setting in which the singularity emerges is less important than the phenomenology of the underlying model, which in some cases resembles that of an equilibrium system with distinct phases and in some cases does not. In this note I want to make a technical point about how dynamical phase transitions are diagnosed, comment on existing classification schemes, and offer some suggestions for how to clarify the nature of systems that display these kinks.\\

\noindent {\em The limiting procedure often used to identify dynamical phase transitions is problematic.} Some authors advocate an approach to dynamical large deviations in which time $T$ is interpreted as if it were a spatial coordinate of a thermodynamic system\c{jack2020ergodicity}. In this approach, SCGF kinks that arise in a joint limit, in which first $T$ and then system size $N$ are taken infinite, are taken as evidence of a first-order dynamical or ``space-time'' phase transition, accompanied by dynamical phase coexistence. I have argued previously that large-deviation singularities are not necessarily accompanied by phase-transition-like behavior: singularities can occur if a system is simply slow, whether or not it supports multiple phases\c{whitelam2018large, whitelam2021varied}. Here I revisit this point in the context of the many-body system of~\cc{buvca2019exact}, which in the double limit displays a singularity that the authors argue indicates a phase transition. I argue instead that that the double-limit procedure produces a breakdown of the large-deviation principle (LDP), from which little can be determined. However, it is possible to take the large-size limit of the model in a way that restores the LDP, and this picture, combined with numerical simulations of the model presented in the paper, does not provide strong evidence for coexisting phases.

The authors of~\cc{buvca2019exact} take the long-time limit in order to obtain the SCGF, $\theta_N(s)$, of a particular time-integrated observable ($s$ is the Legendre-transform parameter), and then take the large-$N$ limit. The result is a kinked $\theta_N(s)$ and a set of large-deviation functions reminiscent of those of the Ising model below its critical temperature, which does display phase coexistence\c{binney1992theory,touchette2009large}. However, several aspects of these functions are inconsistent with known features of the model.  

Principally, the model has a well-defined mean or typical value of the chosen observable, Eq.~(8) of the paper, even in the limit $N \to \infty$. However, the function $\lim_{N \to \infty} N^{-1} \theta_N(s)$, Eq.~(10) of the paper, does not know this: it is not differentiable at the origin, and contains none of the model parameters $\alpha$ to $\delta$ that appear in Eq.~(8). The function $\lim_{N \to \infty} N^{-1} \theta_N(s)$ cannot generate the first cumulant, and so therefore cannot be the cumulant-generating function. Equivalently, the rate function associated with Eq.~(10) possesses a line of zeros. If a large deviation principle applies, we would expect a rate function with a unique zero that corresponds to the model's typical behavior\c{den2008large,touchette2009large}. 

I argue that this model {\em is} described by a large-deviation principle, even in the limit $N \to \infty$, but on a different scale to the one examined in the paper. We can look for a large-deviation principle with large parameter (or speed) $\tilde{T} \equiv T/\tau(N)$, where $\tau(N)$ is the dominant model timescale, by writing $P_T \sim \exp(-T \varphi_N(x)) \equiv \exp(-\tilde{T} \tilde{\varphi}_N(x))$, where $P_T$ is the probability of observing a particular value $x$ of te observable. The rate function on speed $\tilde{T}$ is $\tilde{\varphi}_N = \tau(N) \varphi_N$. The corresponding SCGF is obtained by sending $s \to s/\tau(N)$, giving the function $\theta_N(s/\tau(N))$. 
\begin{figure}[] 
   \centering
   \includegraphics[width=\linewidth]{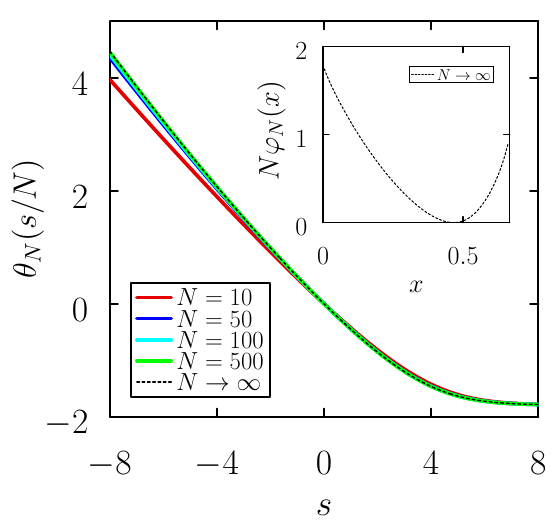} 
   \caption{Large-deviation functions of the boundary-driven cellular automaton of \cc{buvca2019exact} on speed $T/N$. The data collapse in the large-$N$ limit indicates the existence of a well-defined large-deviation principle. Inset: limiting form of the corresponding rate function, which quantifies the logarithmic probability of observing a value $x$ of the dynamical observable studied in that paper. The unique zero corresponds to the model's typical behavior.}
   \label{fig1}
\end{figure}

Given that the model's dynamics is ballistic, I assume $\tau(N)=N$, and plot in \f{fig1} the large-deviation functions on speed $\tilde{T}=T/N$. As $N$ becomes large, these functions tend to a limiting, non-singular form. The colored lines were produced by numerically evaluating Eq.~(4) of~\cc{buvca2019exact}. The black dashed line was produced by taking the limit $N \to \infty$ analytically (having first sent $s \to s/N$) and then evaluating the resulting equation numerically. The resulting function is the limiting form to which the SCGF tends. The inset of \f{fig1} shows the corresponding limiting rate function, $\lim_{N \to \infty} \tilde{\varphi}_N(x)$, which quantifies the logarithmic probability of observing certain values $x$ of the time-integrated observable. As expected from a well-defined large-deviation principle, this has a unique zero corresponding to the typical value of the observable, Eq.~(8) of the paper. That is, the limiting analytic procedure consistent with a well-defined large-deviation principle is to take $T \to \infty$, then rescale the Legendre-transform parameter $s$ by the system's dominant timescale, and then take $N \to \infty$. Similar considerations apply to a random-walk model in the large-size limit\c{whitelam2021varied}, and to finite-size models in the small-rate limit\c{whitelam2018large}.

Does this model exhibit phase coexistence? One way a system could be said to exhibit coexisting dynamical phases is if the probability of observing trajectories of distinct values of a time-integrated observable are equal and special in some way (see e.g. the right-hand panel of Fig. 6 of~\cc{whitelam2021varied}). The limiting functions displayed in~\cc{buvca2019exact} suggest that dynamical trajectories corresponding to large or small values of the chosen observable become increasingly prominent or important as the system is made larger, at least by analogy with the Ising model. My reservation with this analogy is that it is based principally on the fact that the LDP breaks down for both models, rather than an assessment of the physics of each. We know from elementary considerations that the Ising model below its critical temperature has two typical behaviors, which can be identified on the scale $\sqrt{N}$ (in 2D)\c{binney1992theory,touchette2009large}, while the dynamical model has one, which can be identified on the scale $T/\tau(N)$. The logarithmic probability of observing trajectories with values $x$ of the observable is $T/\tau(N)$ multiplied by the function shown in the inset of \f{fig1}, which resembles that of a Poisson-like process with a large timescale, and does not support the idea of coexisting values of $x$.

Another way a system could be said to exhibit coexistence is if the trajectories that realize particular values of the time-integrated observable are intermittent, interconverting between different values of the instantaneous quantity that comprises the time-integrated observable (see e.g. the middle panel of Fig. 6 of~\cc{whitelam2021varied}). Whether or not this is the case cannot be deduced directly from the rate function~\footnote{A linear nonvanishing rate function is suggestive of intermittency, although linear rate functions can also arise in other contexts\c{touchette2009large}.}, and must be assessed by direct calculation of those trajectories. The authors of~\cc{buvca2019exact} rule out this scenario in the text, and show in Fig. 3 some rare trajectories of the model; these are not intermittent.

To summarize, kinked large-deviation functions that emerge in the double-limit procedure of~\cc{jack2020ergodicity} are not necessarily diagnostic of phase coexistence, and are caused for the model of~\cc{buvca2019exact} by a breakdown of the large-deviation principle. The idea of phase coexistence in this model is not supported by the large-deviation functions that result from a limiting procedure modified to restore the LDP, or by the nature of the rare dynamical trajectories calculated in~\cc{buvca2019exact}. I would argue that the evidence points toward this model having dynamics that slows with increasing system size, not cooperative behavior reminiscent of a phase transition.\\

\noindent {\em Existing classification schemes tend to focus on how the singularity emerges, rather than the underlying model phenomenology.} \f{fig2} shows some examples of systems described in the literature as having dynamical phase transitions. It is incomplete (e.g. some systems show singularities away from any limit\c{nyawo2017minimal}), but captures some of the more common types of behavior. Although each system is described as a phase transition by the authors who study it, because of the presence of a large-deviation singularity, there is not consensus among this group as to whether application of the label should depend upon the limit in which the singularity arises. The label is more likely to be applied to systems as we move downwards on~\f{fig2}, because equilibrium phase transitions occur in the large-system-size limit of many-body systems. 

Aside from issues of consistency, one problem with classification-by-singularity is that it is insensitive to phenomenology: the label ``phase transition'' can then be applied to models that are simply slow, and display no features that could be said to resemble distinct phases. The label can also be applied to systems with intermittent conditioned trajectories, which are more phase-transition-like, their behavior in the long-time limit resembling the static behavior of the 1D Ising model at nonzero temperature (which, however, does not exhibit true phase coexistence in equilibrium terms). There also exist models with cooperative dynamics, much as the Ising model has cooperative interactions. These models display large-deviation singularities and phase-transition-like phenomenology that includes a change of typical behavior with a change of model parameters, and a breaking of ergodicity accompanied by the emergence of multiple dynamical phases\c{klymko2017similarity,whitelam2021varied}. However, some authors argue that such behavior should not be called a phase transition\c{jack2019large}, and so it is possible to find literature definitions of ``dynamical phase transition'' that are essentially anticorrelated with the existence of phase-transition-like phenomenology. 

My view is that a more physically meaningful way to classify systems that display large-deviation singularities is by phenomenology. All are different to equilibrium systems in some respects, but we can compare the ways in which their dynamics is similar to the cooperative behavior of an equilibrium system that displays a phase transition. By this measure the degree to which models exhibit phase-transition-like behavior increases from left to right in \f{fig2} (excepting the first column, which indicates systems yet to be classified). The precise dividing line between what we call a dynamical phase transition and what we do not is a matter of interpretation, but in my view should be based on the physics of the model under study.
\begin{figure}[] 
   \centering
   \includegraphics[width=\linewidth]{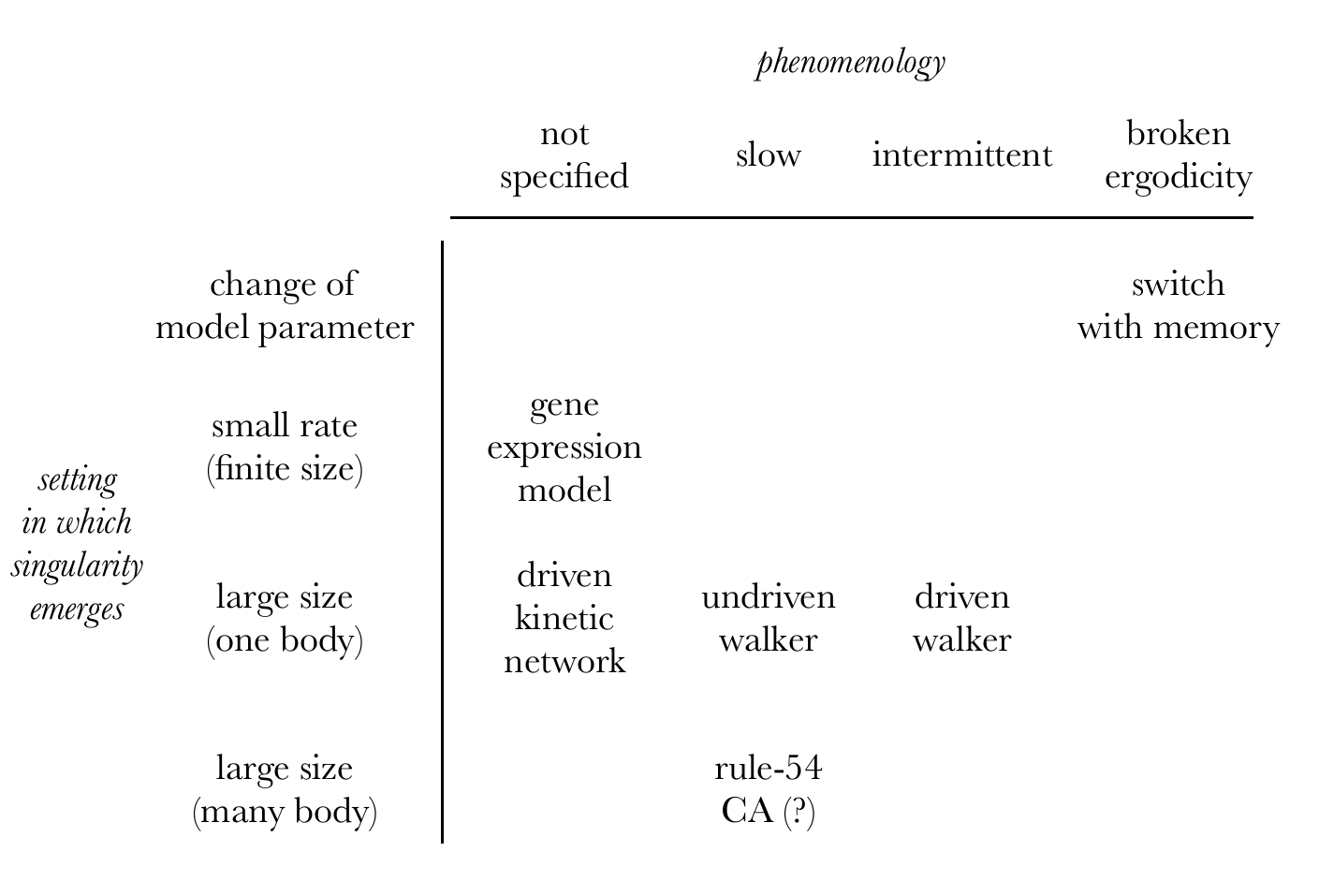} 
  \caption{\label{fig2} Examples of models that exhibit large-deviation singularities: a model of gene expression\c{horowitz2017stochastic}, a driven kinetic network\c{vaikuntanathan2014dynamic}, undriven\c{whitelam2021varied}- and driven\c{nemoto2017finite,whitelam2021varied} random walkers, a cellular automaton\c{buvca2019exact}, and a switch with memory (or growth model)\c{klymko2017similarity,whitelam2021varied}. All are described in the papers indicated as displaying phase transitions, but consensus between these groups of authors does not exist.}
\end{figure}

Absent a consistent classification scheme, one point that could be better clarified in the literature is the meaning of a kinked large-deviation function. If singular behavior is diagnosed in some limit, is this limit consistent with known features of the model, such as its typical behavior? Does the LDP break down entirely, or is there a well-defined LDP on some speed or speeds? Does the kink signal the emergence of distinct trajectory types whose probabilities are equal, or intermittency within a trajectory, or slow dynamics, or some other type of behavior? If the conditioned dynamics of the model is intermittent, what are the dynamical ``phases'' that contribute to the intermittency? Why on physical grounds does the system adopt this behavior in order to achieve, in the least unlikely way, a particular value of a time-integrated observable (see the discussion of Sec. IV of~\cc{whitelam2021varied})?\\

\noindent {\em Acknowledgment.} This work was performed at the Molecular Foundry, Lawrence Berkeley National Laboratory, supported by the Office of Science, Office of Basic Energy Sciences, of the U.S. Department of Energy under Contract No. DE-AC02--05CH11231. 

%

\end{document}